\documentclass[a4paper]{JHEP3}
\usepackage{epsfig}
\usepackage{graphicx}
\usepackage{amsmath} 
\usepackage{amsfonts}
\usepackage{amssymb}
\usepackage{latexsym}

\newcommand{\be}{\begin{equation}}
\newcommand{\ee}{\end{equation}}
\newcommand{\bea}{\begin{eqnarray}}
\newcommand{\eea}{\end{eqnarray}}

\title{ Bell-like examples for spin-1 hidden variable theories}
\author{Kaushik Borah\footnote{ugkaushik@ug.iisc.in, kaushikborah99@yahoo.com}\\
Indian Institute of Science (IISc), Bangalore-560012, India.}
\author{N.D. Hari Dass \footnote{dass@tifrh.res.in}
\\  Visiting Professor, TIFR-TCIS, Hyderabad-500075, India.\\
}

\abstract{
Though John Bell had claimed that his spin-$\frac{1}{2}$ example of a hidden-variable theory(HV) is an \emph{explicit} counterexample 
to von Neumann's proof of the non-existence of hidden variable theories empirically equivalent to quantum mechanics, such examples
 can be so construed only if they met all of von Neumann's requirements. In particular, that they reproduced all the observed predictions 
of quantum theory. To shed light on these aspects, we have, on the one hand, simplified and critically examined Bell's original example and 
on the other hand, constructed explicit such examples for spin-1 systems. We have clarified the relation of our example to the Kochen-Specker 
and Bell's powerful earlier results. Our spin-1 examples are manifestly non-contextual, yet violating the K-S constraints configuration by 
configuration. Nevertheless, they reproduce the correct quantum expectation values and variances for arbitrary linear combinations of 
the beables in one case, and for close approximants to the beables representing K-S constraints. In conformity with the K-S theorem, 
the variance of the K-S constraint is nonzero. The implications of this non-vanishing variance are analysed in detail. In the other example,
we show how this variance can be made arbitrarily small but not zero.  
}


\begin{document}

\maketitle

\section{Introduction}
Quantum mechanics is a highly succesful description of nature over a mind-boggling range of scales from atoms to elementary particles. It is philosophically intriguing and mathematically beautiful and powerful. Yet, all that seems to be coming at the price of 
manifestly non-causal behaviour. This non-causal behaviour of quantum mechanics is at the core of its radically different views
of reality compared to classical physics. It has also necessitated an apparently inherent probabilistic description of nature.

Questions have been asked ever since the early days of Quantum Theory, continuing to the present day, whether this non-causal,
probabilistic nature of quantum theory is merely a facade to a deeper level of description which is causal and deterministic. Going
by the generic tag of \emph{Hidden variable Theories}, hereafter abbreviated as HV, this attempted deeper level of description
posits that there are degrees of freedom that have till now remained hidden from our experience, and their ignorance is what
simulates the randomness associated with quantum phenomena. In this respect HV theories can be likened to statistical mechanics
where ignorance of particle positions and momenta necessitated by the extremely large number of degrees of freedom results in
a statistical or probabilistic description. In quantum theory, the probability description is intrinsic and as per HV theories, the
true description of nature should be sought in terms of HV theories which are causal and deterministic. But some of the degrees of 
freedom of the HV theories are deemed to be unobservable, at least at present. Therefore outcomes of measurements can not reveal
the values of such hidden variables and the fully causal and deterministic laws of HV theories acquire a probabilistic
character. An excellent introduction to the motivations behind HV theories and their expected properties can be found in
von Neumann's classic {\it The Mathematical Foundations of Quantum Theory} \cite{jVNbook}.

But the mathematical foundations of quantum mechanics are so rigid that it appears to be practically impossible for a causal, deterministic and 
classical theory to embrace all its features, though there are a number of recent developments hinting at the possibility of realising
quantum-probability like features in deterministic systems. So much so that von Neumann, in the same book, claimed that a proof of non-existence
of HV theories can be given out of \emph{purely mathematical considerations}. He did give such a proof in his book.

There are two aspects to von Neumann's proof for `No Hidden Variable Theories'. First part is the proof of the absence of what 
he calls \emph{dispersion free ensembles}. This has been severely criticized by Bell on the basis that assumptions in the proof imply 
the absurdity that  the eigenvalue of the sum of two operators always equals the sum of eigenvalues of the individual  operators. 
A careful reading of von Neumann's proof does reveal this weakness, as a result of which the whole proof breaks down. One of us
has attempted to take a relook at this proof without recourse to the offensive assumption \cite{ndhhvproof}. It is intriguing how one of 
the greatest mathematicians could have made such an elementary error. In the course of his proof, von Neumann makes use of a Hilbert space
formalism for the Hidden variable theories too. In fact, he and Koopman had pioneered the use of Hilbert space formalism for
classical systems too.But classical and quantum systems are clearly dramatically different in their physics and in our opinion, the error
of von Neumann was in having overlooked some of these differences at crucial junctures of his proof.

Not only did Bell criticise von Neumann's proof, in our view needlessly harshly, he gave an explicit example for spin-1/2 systems, which
he claimed was with the aim of showing that {\it at the level considered by von Neumann, such an interpretation is not excluded}. Here
Bell meant by 'such an interpretation' a description based on hidden variables. Subsequently, this example of Bell has been elevated to
a \emph{counter-example} to von Neumann's claims.

Our main aim in this paper is to critically examine this example to see whether it can really be construed as a satisfactory counterexample
to von Neumann's proof. What von Neumann sought to prove(though one part of his proof itself was flawed) was the impossibility of a 
hidden variable description that would succesfully 
reproduce all the features of quantum mechanics, for all physical systems. This would be a minimal test for all hidden variable descriptions. 
What Bell's example
succeeded in doing was reproduce the quantum mechanical expectations of linear combinations of the basic observables of a spin-1/2 system.
But 
central to von Neumann's theorem is the search for a hidden variable description that would be succesful in reproducing the \emph{outcomes}
and \emph{expectation values} of {\bf all} the observables for every system. The spin-1/2 system is too simple to address the issues in their full generality.
This follows from the fact that on the quantum mechanical side, the algebra of observables is rather simple. Put another way, it suffices to get the expectation values of linear combinations of the basic observables right to get their, for example, variances correctly too. But already
in the spin-1 case the situation is considerably more complicated; correctly accounting for the expectation values of linear combinations
of observables is no longer adequate to get their variances correctly. One of our aims was therefore to produce a Bell-like example that 
would correctly account for both averages and variances of linear combinations of beables, and we did succeed not only in finding such
an example, but in giving the general prescription for finding such examples.

But any success in producing such examples can not sit well with the Bell-Kochen-Specker theorems, so something must go wrong in their ability
to predict all aspects of quantum mechanics. And we indeed find that examples of first kind fail to reproduce correctly the variance in 
the K-S operator. In quantum mechanics, this operator being a multiple of identity is trivially dispersion free. This is significant because
these Bell-type examples for spin-1 while achieving what Bell's example did for spin-1/2, namely,correctly reproduce the expectation values for arbitrary linear combinations of the 'basic' observables, and even more in the sense of correctly reproducing the variances for such linear
combination of operators, fail nevertheless to reproduce the correct expectation values for more complicated observables like the square of the
K-S constraint. In hidden variable theories, such squares are not necessarily trivial valued. Therefore, Bell's spin-1/2 example can hardly
be taken to imply a counter-example to the claims of von Neumann. 

Nevertheless,we are able to construct hidden variable models for spin-1 such that the K-S operator is arbitrarily dispersion free but never
exactly dispersion free. That this dispersion can never be made exactly zero is the essential content of the Kochen-Specker theorem. While 
their proof is essentially geometric, our analysis adds a state-dependent dimension to it and is to be thought of as complimentary.

The second aspect of von Neumann's proof is based on the absence of what he calls homogeneous ensembles, and indeed, all hidden variable
theories must necessarily violate this criterion if they are to be consistent with quantum mechanics. Surprisingly, this aspect has not
been commented upon, even by Bell, Mermin etc. We show, with Bell's own example, that while this objection against all hidden variable theories is always 
correct, it is empirically very difficult to test.

\section{Quantum Theory \textit{vs} Hidden Variable Theories}
In this section we make a comparison, at very general structural level, between quantum theory and hidden variable theories. We shall also
try to characterise what we consider to be features of an acceptable hidden variable theory.In both of them we need to consider states, 
evolution of states and the results(outcomes) of measurements. The broadest definition of states in any theory is the one given by 
Dirac \cite{diracstate}: {\it States are the embodiment of the collection of all possible measurement outcomes}. It should be appreciated that 
this broad characterisation of states holds equally well in a probabilistic theory like quantum mechanics as well as in deterministic hidden
variable theories. In quantum theory,the outcomes of measurements are generically random but the averages of outcomes(for some observable) are
completely determined by the state. Turning this around, the averages of outcomes for the {\it complete set} of observables determines the 
state.However, the individual outcomes are the eigenvalues of the observable, occurring randomly. This means on the one hand that unlike in
classical systems, no definite value can be associated for observables in a state, and on the other hand, no event by event causal description
is available in quantum theory. The other distinctive feature of quantum theory is that the algebra of observables can be 
{\it non-commutative}. All these features are formally expressed by associating the so called {\it Pure} states by rays in a Hilbert space 
and the observables by {\it self-adjoint operators} acting on the same Hilbert space. The rules of quantum measurement(to be precise, only
for the so called projective measurements) are that 
the outcomes of an observable O are its eigenvalues $o_i$ randomly and associated with the outcome $o_i$ the state changes to the eigenvector
$|o_i\rangle$. The more general description of a quantum state is provided by {\it Density Matrices}. The density matrix corresponding to a
pure state $|\psi\rangle$ is $\rho_\psi\equiv\,|\psi\rangle\langle\psi|$. The state after a measurement can not in general be described by
a pure state and one needs {\it mixed states} for their description. In the example of an observable O measured in a pure state $|\psi\rangle$,
the post measurement mixed state is $\rho_>\,=\,\sum_i\,|\langle\psi|o_i\rangle|^2\,|o_i\rangle\langle o_i|$. The density matrices, for both
pure and mixed states, satisy $tr\,\rho=1,\rho^\dag=\rho$. Only the pure states satisfy $\rho^2=\rho$. It is clear that for a complex 
D-dimensional system, the general density matrix is DxD Hermitian, thus characterised by $D^2-1$ real parameters while pure density matrices
are parametrised by $2D-2$ real parameters.   

The expectation value of the observable O in the pure state $|\psi\rangle$ is given by
\be
\label{psiaverage}
\langle\,O\,\rangle_\psi\,=\,\langle\,\psi\,O\,\psi\rangle
\ee
In terms of the pure density matrix $\rho_\psi\,=\,|\psi\rangle\langle\psi|$ this can be rewritten as
\be
\label{rhopsiaverage}
\langle\,O\,\rangle_\psi\,=\,tr \rho_\psi\,O
\ee
The advantage of the second representation is that it continues to hold for mixed density matrices also i.e
\be
\label{rhoaverage}
\langle\,O\,\rangle_\rho\,=\,tr \rho\,O
\ee
These equations subsume the second important aspect of projective measurements, namely, the Born probability rule according to which the 
frequency with which the outcome $o_i$ occurs in an ensemble measurement is given by $|\langle \psi|o_i\rangle|^2$.


The hidden variable theories on the other hand are to be such that they reproduce {\bf all} the observable consequences of quantum mechanics
without recourse to any randomness. They are supposed to provide a causal event by event description. The randomness of quantum mechanics
is supposed to arise from the lack of information regarding what are called {\it hidden} variables. If the values of these hidden variables
were known, the outcomes of measurements would be completely known. The probability distributions encountered in the quantum descriptions
are attributed to {\it classical} probability distributions for these hidden variables.

In such theories, measurable attributes which correspond to the observables of quantum theory, sometimes called {\it beables} are
supposed to have definite values in a given hidden variable state irrespective of whether they are measured or not. Furthermore, even beables
corresponding to noncommuting observables in quantum mechanics like $S_x,S_y$ are supposed to have definite values in every hidden variable
state. 
An important question is
whether upon measurement the outcome is simply the value of the beable in the given state; another important question is whether the hidden variable state changes when a measurement is performed or not. In conventional classical physics, the answer to the first is in the affirmative
and the answer to the second is in the negative, at least in principle. Though most assertions in quantum theory are probabilistic and hence
only ensemble measurements are meaningful, there are distict features of quantum theory even in measurements on single copies.
Consider the following sequence of measurements on a {\it single copy}, not an ensemble. First measure $S_x$ with outcome spin-up in 
x-direction. 
Next, consider a measurement of $S_y$ and let the outcome be spin-down in y-direction.
At this point a hidden variable theory in which measurement does not change the hidden variable state(we shall come back to what
states in a HV theory are later) can be compatible with quantum theory. While in quantum theory the interpretation for the above is that
irrespective of the initial state, the state after the first measurement is $|\uparrow\rangle_x$ associated with the random outcome $S_x=+$,
the state after the second measurement is $|-\rangle_y$ associated with the random outcome $S_y=-$. A hidden variable theory in which the
act of measurement has no influence on the hidden variable state, the explanation would be that in the given hidden variable state the value
of the beable $S_x$ is +, while the value of $S_y$ in the same hidden variable state is -. Now let us perform a third measurement on the same 
copy and let this be a measurement of $S_x$. According to quantum mechanics, the outcome is again {\it random} with a 50\% chance of getting 
either + or -. On the other hand, in a hidden variable theory where the state does not change due to measurements, the third measurement must
always yield the same outcome as the first measurement, thus in disagreement with the predictions of quantum mechanics. The only way out is
that measurements must change the hidden variable state. But will that be in conflict with the deterministic nature of hidden variable 
theories? Not necessarily as long as the change in state can be described deterministically i.e given the initial state and the beable that is
measured, the state after measurement must be determined uniquely. In quantum mechanics the state change is random.
The above example brings out the very essence of noncommutativity in quantum mechanics and we see that the only way for hidden variable 
theories whose algebra of beables has to be necessarily commuting to be compatible with quantum theory is for measurements to affect the
state albeit in a deterministic way.

A hidden variable state is completely described by a set of values of all the so called \emph{Beables} which are
like classical observables, supplemented by some hidden variables.
As hidden variable theories are fully deterministic by construction, the apparent randomness arises due to the 
statistical distribution of hidden variables (say $\rho(\lambda)$). When we measure a beable (say, $\textbf{B}$ ) in HV, the outcome of 
the measurement picks up a value for a particular value of the corresponding hidden variables. Then to calculate the expectation value of 
the beable we have to integrate the outcome of the beable $\textbf{B}_{oc}$ over the distributions of the hidden variables $\rho(\lambda)$ 
i.e., $\langle \textbf{B} \rangle = \int \textbf{B}_{oc} \rho(\lambda) d\lambda$.

\section{Some mathematical preliminaries}
In this section we prove some simple, but very useful results for our general analysis. We consider sign functions depending on some
random variable $\lambda$(later on one of the hidden variables)
which obeys a normalized power law distribution function $\rho(\lambda)$: 
\be 
\label{eq:hvdistrib}
\rho_n(\lambda) = N_n\,(2n+1)\,\lambda^{2n} 
\ee
where $N_n$ is a normalisation constant. This family of distributions for the hidden variable $\lambda$ is parametrised by the integer n
taking values $n\,=\,0,1,2,\ldots$. The range of the hidden variable is taken to be 
$ |\lambda| < S_n^\frac{1}{2n+1}$.
The normalization condition is given by,
\be
\label{eq:hvdistribnorm}
\int \rho_n(\lambda) d\lambda = 2N_n\,S_n\,= 1\,\rightarrow\,S_n\,=\,\frac{1}{2N_n}  
\ee
Now we calculate the expectation value of the $sign$ function,
\be 
\label{eq:hvsgn}
\chi_n(\lambda,\xi)= sign(\lambda + |\frac{\xi}{2N(n)}|^\frac{1}{2n+1})
\ee
where $|\xi|\,\le\,1$. 
Thus,
\bea 
\label{eq:hvsgnav}
\int \rho_n(\lambda) \chi_n(\lambda,\xi) d\lambda &=& N(n)(2n+1)[-\int_{-S_n^\frac{1}{2n+1}}^{-|\frac{\xi}{2N(n)}|^\frac{1}{2n+1}}+\int_{-|\frac{\xi}{2N(n)}|^\frac{1}{2n+1}}^{S_n^\frac{1}{2n+1}}]\lambda^{2n}d\lambda \nonumber\\ &=& |\xi|  
\eea
Hence, 
\be
\label{eq:hvsgnav2}
\int \rho_n(\lambda) \chi_n(\lambda,\xi) d\lambda = |\xi|
\ee
Remarkably, the averages, do not depend on either n or $S_n$! 
The simplest case of $n=0$ is the flat distribution used by Bell in his
original example. For that particular case $N_0=1,S_0=\frac{1}{2}$. 

Next, we consider the two \textit{sign} functions:

\be 
\chi_n(\lambda,\xi_1)= sign(\lambda + |\frac{\xi_1}{2N(n)}|^\frac{1}{2n+1})sign(\xi_1)
\ee

\be 
\chi_n(\lambda,\xi_2)= sign(\lambda + |\frac{\xi_2}{2N(n)}|^\frac{1}{2n+1})sign(\xi_2)
\ee
where $|\xi_1|\le\,1$ and $|\xi_2|\le\,1$.

And we calculate the average of the product of the two sign functions:
\bea
\langle \chi_n(\lambda,\xi_1) \chi_n(\lambda,\xi_2) \rangle &=& \int \rho_n(\lambda) \chi_n(\lambda,\xi_1) \chi_n(\lambda,\xi_2) d\lambda \nonumber\\ 
&=& N_n (2n+1) \int_{-S_n^{\frac{1}{2n+1}}}^{S_n^{\frac{1}{2n+1}}}\lambda^{2n} \chi_n(\lambda,\xi_1) \chi_n(\lambda,\xi_2) \nonumber\\ &=& N_n \{[\lambda^{2n+1}]_{-S_n^{\frac{1}{2n+1}}}^{-\frac{1}{2}|\xi_1|^{\frac{1}{2n+1}}} - [\lambda^{2n+1}]_{-\frac{1}{2}|\xi_1|^{\frac{1}{2n+1}}}^{-\frac{1}{2}|\xi_2|^{\frac{1}{2n+1}}} + [\lambda^{2n+1}]_{-\frac{1}{2}|\xi_2|^{\frac{1}{2n+1}}}^{S_n^{\frac{1}{2n+1}}} \} \nonumber\\ &=& 2S_nN_n + |\xi_2| - |\xi_1| \nonumber\\ &=& 1 + |\xi_2| - |\xi_1|
\eea

There is no $S_n$ or n dependence even in these expectation values. 
  
\section{Bell's spin-$\frac{1}{2}$ example and our modification}
Now we discuss the famous {\it counterexample} to von Neumann first discussed by Bell. He constructed an explicit example for a spin-1/2
system such that for beables corresponding to ${\vec\beta}\cdot{\vec\sigma}$, the hidden variable model with a single real
$\lambda$ uniformly distributed in the interval $|\lambda|\,\le\,1/2$, the expectaion values in the quantum mechanical state 
$|\psi\rangle_0 = \begin{pmatrix}
    1\\
    0
  \end{pmatrix}$ 
are correctly reproduced.
  
Bell chooses
\be 
\label{eq:Bellexample}
\langle\,{\vec \beta}\cdot{\vec S}\,\rangle_{OC} = |\vec{\beta}|sign(\lambda|\vec{\beta}|+\frac{1}{2}|\beta_z|)signX
\ee
as the deterministic formula for the outcome of the particular beable. 
In the above,  
\bea
\label{eq:betaspl}
 X &=& \beta_z ~~ if~ \beta_z \neq 0 \nonumber\\ &=& \beta_x ~~ if~ \beta_z = 0, \beta_x \neq 0 \nonumber\\&=& \beta_y ~~ if~ \beta_z =0, \beta_x =0  
\eea
and
\bea
signX &=& +1 ~~ if~ X \geq 0 \nonumber\\ &=& -1 ~~ if~ X < 0 
\eea 
It is easy to verify that for every $\lambda$, the outcome is either $|{\vec \beta}|$ or its negative, and that the average value of
${\vec\beta}\cdot{\vec\sigma}$ is indeed $\beta_z$.
In this example we see the minimum requirements of an acceptable hidden variable theory; it should be in conformity with all the observed predictions of quantum mechanics. One such are the event by event outcomes, and the other are the frequencies with which these outcomes occur. If we
denote by $P_\pm$ the probabilities of the outcomes $\pm$,
\be
\label{bellaverage}
\langle\,{\vec \beta}\cdot{\vec \sigma}\,\rangle_\psi\,=\,(P_+-P_-)\rightarrow P_++P_-=1
\ee
Therefore predicting the correct average is also tantamount to predicting the correct probabilities. But for more complicated quantum systems
this is certainly not true.

We shall recast this example differently to bring out its interpretaion more clearly. As stated, the reason for the expectation value to be
$\beta_z$ is obscure as $\beta_z$ is playing many roles in the example. One is that 
$\beta_z = \langle\psi_0|{\vec \beta}\cdot{\vec \sigma}|\psi_0\rangle$; the other is the special role it plays in eqn.(\ref{eq:betaspl}).
We shall instead work with 
a general quantum mechanical state $|\psi\rangle = \begin{pmatrix}
    \alpha^\prime\\
    \beta^\prime
  \end{pmatrix}$. 
Now consider the beable ${\vec\beta}\cdot{\vec S}$ where $S_i$ are the beables that correspond to the quantum observables $\sigma_i$. 
The quantum outcomes are still $\pm\,|{\vec \beta}|$, while the
quantum expectation value of ${\vec\beta}\cdot{\vec\sigma}$ is ${\vec\beta}\cdot{\vec\epsilon}$ where
\be
\label{qexp}
{\vec\epsilon}\,=\,\langle\psi|{\vec\sigma}|\psi\rangle
\ee
On the hidden variable side, the outcome rule
\be
\label{modbelloc}
({\vec\beta}\cdot{\vec S})_{OC}\,=\,|{\vec\beta}|\,sign({\vec\beta}\cdot{\vec\epsilon})\,\chi()
\ee
where $\chi$ is a sign function that depends on the hidden variable state, reproduces the expectation provided
\be
\label{modbellcond}
\langle\,\chi\,\rangle_\lambda\,=\,\frac{{\vec\beta}\cdot{\vec\epsilon}}{|{\vec\beta}|}
\ee
On using the mathematical preliminaries of Sec.3, this gives, when the hidden variable $\lambda$ is uniformly distributed,
\be
\label{modebellchi}
\chi(\lambda,{\vec\epsilon})\,=\,sign(\lambda\,+\,\frac{{\vec\beta}\cdot{\vec\epsilon}}{2|{\vec\beta}|})
\ee
Thus our modification achieves the same results as the Bell example without complicated conditions like those of eqn(\ref{eq:betaspl}). It
also brings out the fact that the hidden variable state in this case is parametrised by the four real parameters $(\lambda,{\vec\epsilon})$.
In particular, $\lambda$ should be treated as the hidden variable.
%
%
The outcomes are explicitly given by
\be
\label{eq:belloutcome2}
|\vec{\beta}|sign({\vec\beta}\cdot{\vec\epsilon})~~~~~~~ when \lambda > -\frac{1}{2|\vec{\beta}|}|{\vec\beta}\cdot{\vec\epsilon}|
\ee
Likewise,
\be
\label{eq:belloutcome2p}
-|\vec{\beta}|sign({\vec\beta}\cdot{\vec\epsilon})~~~~~~~ when \lambda < -\frac{1}{2|\vec{\beta}|}|{\vec\beta}\cdot{\vec\epsilon}|
\ee


It should be recalled that  the $\epsilon_i$'s label the states(completely) on the quantum mechanical side. On the hidden variable side, the states
need in addition the $\lambda$ for their specification.
\subsection{Is this a satisfactory counterexample?}
Can the Bell example be taken as a counterexample to the so called von Neumann theorem in its entirety? It is true that von Neumann's
arguments were flawed and that he had assumed an important ingredient concerning the eigenvalues of the sum of noncommutating observables
on the basis of which he had concluded the impossibility of a hidden variable description(as remarked earlier, this was only in one part of
the proof). To that aspect, the Bell example is indeed a concrete counterexample. But von Neumann sought something much deeper. He wanted 
acceptable hidden variable theories to explain {\bf all} succesful predictions of quantum mechanics. The spirit of his proof was that the
mathematical structures of the two have to be so different that it will be impossible to reconcile the two. This is the reason that he asserted
that a proof of the impossibility of hidden variable theories can be given on a purely mathematical basis.

It is clear from his detailed and careful analysis of hidden variable theories that it was the radical difference in the nature of the algebra
of observables that he was most concerned about. In particular the challenge was in uncovering the noncommutative essence of quantum theory
in a fundamentally commuting hidden variable theory. One aspect of this was of course the additivity of eigenvalues issue which was messed up
in von Neumann's analysis. It is pertinent to remark here that it is not as if one of the world's greatest mathematicians was unaware of this
elementary property of matrices! However, this is only one of the aspects of noncommutativity in quantum mechanics. Among the others is what we have already discussed i.e the outcomes of repeated measurements on a single copy. We argued that unless measurements change the hidden variable
state, albeit deterministically, this aspect of quantum mechanics can not be reconciled. In its fullest generality, the sequence of measuremens
can involve not just $S_x,S_y,S_z$ but beables in arbitrary directions like ${\vec S}\cdot{\vec n}_1,{\vec S}\cdot{\vec n}_2...$. Not only
the details of the sequence of outcomes on single copies, but the sequence of probability distributions should also be in agreement with what 
quantum theory ordains. Unless a
deterministic rule for change of states is given which can consistently describe this, one can not accept the Bell example as providing a
counterexample to von Neumann's thinking which should really be seen in broader terms as the impossibility of providing a hidden variable
description for all aspects of quantum mechanics.

One of the powers of the noncommutative algebra of observables in quantum mechanics is that expectation values of the products of strings
of operators can be reduced to simpler expectation values. On the hidden variable side, any product, not necessarily commutators
should also be beables and unless one invokes measurement aspects, expectation values of products of beables can not be made to have any
bearing on the expectations of commutators and anticommutators of observables in quantum mechanics. The way out is certainly not working
them out first using the quantum algebras! The hidden variable theories can only invoke the empirical aspects and not the detailed mathematical
structures of quantum mechanics unless the latter can sit comfortably in their frameworks.
\section{Our prescription of hidden variable formula for spin-1 particle}
\subsection{General case for spin-1}
Now we consider hidden variable theories for spin-1. On the quantum side, this is a much richer system. Firstly, we have two nontrivial
mutually commuting observables. The algebra of observables is such that squares of observables are not trivial, like the identity operator,
any more. This means that variances and means are independent, and the hidden variable theory has to account for both.
A first reaction is that because of 
the Bell-Kochen-Specker theorem one can not extend Bell's example to spin - 1. It is of course true that the Bell-Kochen-Specker theorem
would not be consistent with a hidden variable theory that would explain {\bf all} aspects of spin-1 quantum mechanics. Nevertheless,
it is worth investigating if at least a Bell-like example that would correctly reproduce the correct averages of linear operators can
be derived. Since now variances are independent of means, one can pose the more ambitious question whether a deterministic Bell-like
formula can be obtained that would simultaneously reproduce the correct averages and variances.
The main result of this work is that such Bell-like examples correctly predicting both averages and variances can indeed be found and we
give a systematic way of constructing them.

Then the question arises as to where exactly the conflict with the Bell-kochen-Specker theorems lies and whether one can quantify this 
disagreement? We show that this is revealed through the dispersion in the K-S constraint
\be
\label{eq:K-Sconstraint}
B\,=\,S_x^2\,+\,S_y^2\,+\,S_z^2
\ee 
While on the quantum mechanical side this is twice the identity and is therefore always dispersion free, on the HV side it can
have nontrivial dispersion. We explicitly show that this is so. Therefore we at once have situations where Bell-like examples can be 
constructed that agree with quantum averages and dispersions of linear operators, they disagree with quantum mechanics for dispersions
of quadratic operators!This explicitly demonstrates the inadequacy of Bell-like examples as true counter-examples to any theorem on
the impossibility of a hidden variable description of nature. Incidentally, since our class of examples correctly account for both averages
and variances of linear combinations of beables, they automatically reproduce the correct average for the K-S constraint i.e
\be
\label{eq:hvksconstraint}
\langle\,(S_x^2+S_y^2+S_z^2)\,\rangle_{HV}\,=\,2
\ee

For spin-1 case the basic quantum mechanical observables are in SU(3) Lie algebra. One could take these to be the 8 Gell-Mann matrices
$\Lambda_i$. We consider operators linear in the $\Lambda_i$ of the type
\be
\label{eq:lambdalinear}
O\,=\, \sum_{i=1}^{8} \beta_i \Lambda_i  
\ee
The $\Lambda_i$ obey the algebra:
\be   
\label{eq:gellmanncomm}
[\Lambda_i,\Lambda_j]=2if_{ijk}\Lambda_k
\ee
and,
\be 
\label{eq:gellmannanti}
\{\Lambda_i,\Lambda_j\}=\frac{4}{3}\delta_{ij}+2d_{ijk}\Lambda_k
\ee

The quantum mechanical outcomes for the operator O
for general state $|\psi\rangle = \begin{pmatrix}
    \alpha^\prime\\
    \beta^\prime\\
    \gamma^\prime
  \end{pmatrix}$ would be the eigenvalues  $\lambda_1, \lambda_2$ and $-(\lambda_1 + \lambda_2)$ as the Gell-Mann matrices are traceless 
$Tr(\Lambda_i)=0$. 

The corresponding beable for the spin - 1 hidden variable theory can be written as ${\cal O}= \sum_{i=1}^{8} \beta_i S_i$. In quantum mechanics, a pure
state of the spin-1 system is parametrised by four real numbers. 
If the state is mixed, one needs 8 real parameters which can be taken to be the expectation values of the
8 Gell-Mann matrices defined via the density matrix according to
\be
\label{eq:hvdensityspin1}
\langle\,\Lambda_i\,\rangle_\rho = Tr\, \rho\,\Lambda_i\,= \epsilon_i
\ee
The advantage of the density matrix approach is that it can also be used both for pure and mixed states. The only difference then is that the 
8 $\epsilon_i$ are actually functions of only 4 real parameters. For our purposes, it is more convenient to label the hidden variable state
by the 8 $\epsilon_i$, and the hidden variables $\mu_i$. 
Now we have three possible outcomes and a single sign function will not suffice, as was the case for qubits. At least two such sign functions
are needed. 
\textbf{Outcome formula for spin-1 case}
We now discuss a systematic way of constructing Bell-like examples for spin-1. As we need at least two independentsign functions, 
the minimum number of hidden variables required is two and let us denote them by $\mu_1, \mu_2$. To keep the discussion transparent, we
take them to be uniformly distributed and hence $|\mu_i|\,\le 1/2$. Though the general hidden variable state is to be parametrised by
8 more of the $\epsilon_i$, the sign functions will be seen to be dependent on fewer number of functions of these 10 parameters. The key
to this observation is that apart from the outcomesi of a quantum measurement, their probabilities exhaust the information content. We shall
demonstrate this explicitly. The quantum outcomes are labelled by $\lambda_1,\lambda_2,\lambda_3$. Since the observables are traceless, they
are constrained by $\lambda_1+\lambda_2+\lambda_3=0$. Let $p_1,p_2,p_3$ be their respective probabilities. Hence $p_1+p_2+p_3=1$. The
expectation values of the observable and its square on the quantum side are given by
\be
\label{eq:qavspin1}
\langle|O|\rangle_\psi = \sum\,\lambda_i\,p_i\quad\quad \langle|O^2|\rangle_\psi = \sum\,\lambda^2\,p_i
\ee

On the hidden variable side, let the corresponding beable be ${\bf B}$. Their outcomes in the hidden variable world must be the same 
$\lambda_i$. The most general outcome on the HV side must be of the form 
\be 
\label{eq:genocspin1}
(B)_{OC} = a + b \chi_1 + c \chi_2 + d \chi_1 \chi_2
\ee
where a,b,c,d are functions of $\lambda_i$ while $\chi_i$ are functions of $\mu_i,\epsilon_i$. This must be such that for various combinations
of $\chi_1,\chi_2 = \pm$, the outcomes are precisely $\lambda_i$. To obtain the general formulae we shall relax the condition that 
$\sum\,\lambda_i=0$ but restore this constraint in actual applications.
%
%
%
%
%

%
%

Since there are four such possibilities it is clear that some $\lambda_i$ must repeat. Let the repeated outcome be labelled $\lambda_1$
and $\lambda_2$,  $\lambda_3$ for the remaining.

In the generic case when $\lambda_2\ne\lambda_3$ and neither of them equals $\lambda_1$, there are $2\cdot^4C_2=12$ different cases to be 
considered.
6 different types are listed below, and the other six can be obtained by interchanging $\lambda_2\leftrightarrow\lambda_3$.

\begin{center}
 \begin{tabular}{||c |c ||c |c |c |c |c | c||} 
 \hline
  $\chi_1$ & $\chi_2$ & $I$ & $II$ & $III$ & $IV$ & $V$ & $VI$ \\ [0.5ex] 
 \hline\hline
 1 & 1 & {$\lambda_1$} & $\lambda_2$ & {$\lambda_1$} & {$\lambda_1$} & $\lambda_2$ & $\lambda_2$ \\ 
 \hline
 1 & -1 & $\lambda_2$ & {$\lambda_1$} & $\lambda_2$ & {$\lambda_1$} & {$\lambda_1$} & $\lambda_3$  \\
 \hline
 -1 & 1 & $\lambda_3$ & {$\lambda_1$} & {$\lambda_1$} & $\lambda_2$ & $\lambda_3$ & {$\lambda_1$}  \\
 \hline
 -1 & -1 & {$\lambda_1$} & $\lambda_3$ & $\lambda_3$ & $\lambda_3$ & {$\lambda_1$} & {$\lambda_1$}  \\[1ex]
 \hline
\end{tabular}
\end{center}

Let us check all the possible cases. We simplify the analysis by restricting to the cases when $\langle\,\chi_1\chi_3\,\rangle\,=\,\langle\,\chi_1\,\rangle\langle\,\chi_2\rangle$. It should be emphasised that this condition of uncorrelated chi's is not required; but our aim here
is to demonstrate that there are large classes of Bell-like examples for spin-1. The analyses given here can always be extended by
dropping this simplifying condition.
%

\textbf{Case-I}:  

In this case, the conditions to be obeyed are:
\be
\label{eq:caseI} 
a + b + c + d = \lambda_1, ~ a - b - c + d = \lambda_1, ~ a + b - c - d = \lambda_2, ~ a - b + c - d = \lambda_3
\ee
After solving these, the outcome formula for this case takes the form
\be
\label{eq:caseIoc}
{(B)}_{OC} = [ \frac{\lambda_1}{2} + \frac{1}{4}(\lambda_2+\lambda_3) ] + \frac{1}{4}(\lambda_2-\lambda_3) \chi_1 - \frac{1}{4}(\lambda_2-\lambda_3) \chi_2 + [ \frac{\lambda_1}{2} - \frac{1}{4}(\lambda_2+\lambda_3) ] \chi_1 \chi_2
\ee
On the other hand, this expectation value is also given by
\be
\label{eq:caseIprob}
\langle {B} \rangle = p_1 \lambda_1 + p_2 \lambda_2 + p_3 \lambda_3
\ee

Now comparing equations (\ref{eq:caseIoc}) and (\ref{eq:caseIprob}) we have,
\be
\label{eq:caseIavchi}
\langle \chi_1 \rangle \langle \chi_2 \rangle = 2p_1 -1
\ee 
and, 
\be
\label{eq:caseIchidiff}
\langle \chi_1 \rangle - \langle \chi_2 \rangle = 2(p_2 - p_3)
\ee  
Now solving for $\langle \chi_2 \rangle$:
\be
\label{eq:caseIchi2}
\langle \chi_2 \rangle = -(p_2-p_3) \pm \sqrt{(p_2-p_3)^2+(2p_1-1)}
\ee
When $2p_1+(p_2-p_3)^2\,<\,1$, one sees that the square root becomes imaginary.  
$p_1 = 0$, $p_2 = p_3 = \frac{1}{2}$ is an example of that. Consequently, $\langle \chi_2 \rangle $ can become imaginary,
and this case has to be rejected. It should however be emphasised that these conclusions were based on the $\chi$'s being uncorrelated.
Relaxing this condition may make this valid but we shall not go into it.

\textbf{Case-II}:  
The functions a,b,c,d for this case satisfy
\be
\label{eq:caseII} 
a + b + c + d = \lambda_2, ~ a - b - c + d = \lambda_3, ~ a + b - c - d = \lambda_1, ~ a - b + c - d = \lambda_1
\ee
The outcome formula for this case is given by:
\be
\label{eq:caseIIoc}
{(B)}_{OC} = [ \frac{\lambda_1}{2} + \frac{1}{4}(\lambda_2+\lambda_3) ] + \frac{1}{4}(\lambda_2-\lambda_3) \chi_1 + \frac{1}{4}(\lambda_2-\lambda_3) \chi_2 + [ -\frac{\lambda_1}{2} + \frac{1}{4}(\lambda_2+\lambda_3) ] \chi_1 \chi_2
\ee
Following what we did for case I, we find 
\be
\label{eq:caseIIchis}
\langle \chi_1 \rangle \langle \chi_2 \rangle = 1 - 2p_1 
\ee 
and, 
\be
\label{eq:caseIIchis2}
\langle \chi_1 \rangle + \langle \chi_2 \rangle = 2(p_2 - p_3)
\ee  
Solving for $\langle \chi_1 \rangle$:
\be
\label{eq:caseIIchi1}
\langle \chi_1 \rangle = (p_2-p_3) \pm \sqrt{(p_2-p_3)^2-(1-2p_1)}
\ee
Once again this case has to be rejected as for $2p_1+(p_2-p_3)^2\,<\,1$, $\chi_1$ can become imaginary.

\textbf{Case-III}:  

The relevant equations for this case are:
\be
\label{eq:caseIII} 
a + b + c + d = \lambda_1, ~ a - b - c + d = \lambda_3, ~ a + b - c - d = \lambda_2, ~ a - b + c - d = \lambda_1
\ee
and,
\be
\label{eq:caseIIIoc}
{(B)}_{OC} = [ \frac{\lambda_1}{2} + \frac{1}{4}(\lambda_2+\lambda_3) ] + \frac{1}{4}(\lambda_2-\lambda_3) \chi_1 + [ \frac{\lambda_1}{2} - \frac{1}{4}(\lambda_2+\lambda_3)] \chi_2 +  \frac{1}{4}(\lambda_3 -\lambda_2) \chi_1 \chi_2
\ee
The $\chi$'s are determined exactly as before:
\be
\label{eq:caseIIIchi2}
 \langle \chi_2 \rangle = 2p_1 - 1 
\ee 
and, 
\be
\label{eq:caseIIIchi1}
\langle \chi_1 \rangle = \frac{p_2 - p_3}{1 - p_1}
\ee  
For all values of $p_i$, the $\chi$'s are bounded correctly: $|\langle \chi_1 \rangle| $, $|\langle \chi_2 \rangle| $ $ < 1 $

An important consistency check is provided by
\be
\label{eq:quadtest}
\langle\,O^2\,\rangle_{HV} = \sum\,p_i\,\lambda_i^2
\ee
At the level of outcomes, we propose that in the HV theory
\be
\label{eq:squareoc}
(O^2)_{OC}\,=\,(O_{OC})^2
\ee
Then we have explicitly verified that eqn.(\ref{eq:quadtest}) holds. We have omitted the details which are tedious. This consistency
requirement guarantees that not only quantum expectation values but also variances of observables like O will be reproduced by the HV theory.
 
Hence \textit{Case-III} is \textbf{possible}.

\textbf{Case-IV}:  

The relevant equations for case IV:
\be
\label{eq:caseIV} 
a + b + c + d = \lambda_1, ~ a - b - c + d = \lambda_3, ~ a + b - c - d = \lambda_1, ~ a - b + c - d = \lambda_2
\ee
and, the outcome formula,
\be
\label{eq:caseIVoc}
{(B)}_{OC} = [ \frac{\lambda_1}{2} + \frac{1}{4}(\lambda_2+\lambda_3) ] + [ \frac{\lambda_1}{2} - \frac{1}{4}(\lambda_2+\lambda_3)] \chi_1 + \frac{1}{4}(\lambda_2-\lambda_3) \chi_2 +  \frac{1}{4}(\lambda_3 -\lambda_2) \chi_1 \chi_2
\ee
The relations between averages of $\chi$'s and probabilities of outcomes: 
\be
\label{eq:caseIVchi1}
 \langle \chi_1 \rangle = 2p_1 - 1 
\ee 
and, 
\be
\label{eq:caseIVchi2}
\langle \chi_2 \rangle = \frac{p_2 - p_3}{1 - p_1}
\ee  
This too is always consistena, and,t  
$|\langle \chi_1 \rangle| $, $|\langle \chi_2 \rangle| $ $ < 1 $
Here too we have carried out the quadratic consistency test but we skip the details.

\textbf{Case-V}:  

Summarising
\be
\label{eq:caseV} 
a + b + c + d = \lambda_2, ~ a - b - c + d = \lambda_1, ~ a + b - c - d = \lambda_1, ~ a - b + c - d = \lambda_3
\ee
and,
\be
\label{eq:caseVoc}
{(B)}_{OC} = [ \frac{\lambda_1}{2} + \frac{1}{4}(\lambda_2+\lambda_3) ] + \frac{1}{4}(\lambda_2-\lambda_3)   \chi_1 + [ -\frac{\lambda_1}{2} + \frac{1}{4}(\lambda_2+\lambda_3)]\chi_2 +  \frac{1}{4}(\lambda_2 -\lambda_3) \chi_1 \chi_2
\ee
The determination of avrage $\chi$'s: 
\be
\label{eq:caseVchi2}
 \langle \chi_2 \rangle = 1 - 2p_1  
\ee 
and,
\be
\label{eq:caseVchi1}
\langle \chi_1 \rangle = \frac{p_2 - p_3}{1 - p_1}
\ee  
$|\langle \chi_1 \rangle| $, $|\langle \chi_2 \rangle| $ $ < 1 $ and case V is fine. It is not necessary to separately verify the quadratic
test for this case as by the redefinitions $\chi_2^\prime = -\chi_2,\chi_1^\prime = \chi_1$, cases III and V can be mapped into each other.

\textbf{Case-VI}:  

Finally, the results for case VI are:
\be
\label{eq:caseVI} 
a + b + c + d = \lambda_2, ~ a - b - c + d = \lambda_1, ~ a + b - c - d = \lambda_3, ~ a - b + c - d = \lambda_1
\ee
and,
\be
\label{eq:caseVIoc}
{(B)}_{OC} = [ \frac{\lambda_1}{2} + \frac{1}{4}(\lambda_2+\lambda_3) ] + [ -\frac{\lambda_1}{2} + \frac{1}{4}(\lambda_2+\lambda_3)] \chi_1 + \frac{1}{4}(\lambda_2-\lambda_3) \chi_2 +  \frac{1}{4}(\lambda_2 -\lambda_3) \chi_1 \chi_2
\ee
The $\chi$'s are determined to be:
\be
\label{eq:caseVIchi1}
 \langle \chi_1 \rangle = 1 - 2p_1  
\ee 
and, 
\be
\label{eq:caseVIchi2}
\langle \chi_2 \rangle = \frac{p_2 - p_3}{1 - p_1}
\ee  
$|\langle \chi_1 \rangle| $, $|\langle \chi_2 \rangle| $ $ < 1 $ and this case is consistent too. This can also be seen from the
redefinitions $\chi_1^\prime = -\chi_1,\chi_2^\prime = \chi_2$ which map case VI to case IV. Hence there is no need to explicitly
verify the quadratic test for case VI.
With the results given here one can analyse the hidden variable prescription for operators of the type eqn.(\ref{eq:lambdalinear}).
One can combine the expressions for $\langle\,\chi_i\,\rangle_{HV}$ with the mathematical preliminaries of Sec.3 to construct the
sign functions themselves and from them the explicit formulae for the outcomes.

Such a general study of spin-1 hidden variable theories is interesting in its own right and will be presented elsewhere. Instead, we wish to
focus on the implications of these type of examples for the Bell-Kochen-Specker theorem (more precisely, the implications of the BKS theorem
for examples of this type!). For this, the Gell-Mann basis is not suitable and instead one ought to work in the so called angular momentum
basis. This is discussed in the next section.

\section{Spin-1 analysis in angular momentum basis and the Kochen-Specker Theorem}

To make the connection of the spin-1 case to the Bell-Kochen-Specker theorem, it is necessary to work in the so called \emph{angular momentum}
basis rather than the Gell-Mann basis. This is because 
\be
\label{eq:GM_KS_constraint}   
\Lambda_1^2 + \Lambda_2^2 + \Lambda_3^2 \ne 2\,I
\ee
where I is the 3x3 unit matrix.On the other hand, the three(always three irrespective of spin value) angular momentum operators
$\Sigma_i$ for spin-1 satisfy
\be
\label{eq:QM_KS_constraint}   
\Sigma_x^2 + \Sigma_y^2 + \Sigma_z^2 = 2\,I
\ee
Therefore, in place of $\Lambda_1,\Lambda_2,\Lambda_3$ we introduce $\Sigma_i$ given below:
$\vec{\Sigma}$ are spin angular momentum matrices in SU(2) sub-algebra, 

\be
\label{eq:Sigmas}
\Sigma_x = \begin{pmatrix}
    0 & \frac{1}{\sqrt{2}} & 0\\
    \frac{1}{\sqrt{2}} & 0 & \frac{1}{\sqrt{2}}\\
    0 & \frac{1}{\sqrt{2}} & 0
  \end{pmatrix},
\Sigma_y = \begin{pmatrix}
    0 & \frac{-i}{\sqrt{2}} & 0\\
    \frac{i}{\sqrt{2}} & 0 & \frac{-i}{\sqrt{2}}\\
    0 & \frac{i}{\sqrt{2}} & 0
  \end{pmatrix},
\Sigma_z = \begin{pmatrix}
    1 & 0 & 0\\
   0 & 0 & 0\\
    0 & 0 & -1
  \end{pmatrix} 
\ee
Nevertheless, for $i=1,..,3$, both $\Sigma_i$ and $T_i\equiv\frac{\Lambda_i}{2}$ satisfy the same commutation relations
\be
\label{eq:angmomalgebra}
[X_i,X_j]\,=\,i\,\epsilon_{ijk}\,X_k
\ee
Now, the quantum mechanical observable ${\vec \beta}\cdot{\vec \Sigma}$ has the eigenvalues $\pm\,|{\vec \beta}|,0$. Our considerations
for the Gell-Mann basis can simply be taken over on recognising:
\be
\label{eq:angmom2gellmann}
\Sigma_1 = \frac{1}{\sqrt{2}}(\Lambda_1+\Lambda_6)\quad\quad \Sigma_2 = \frac{1}{\sqrt{2}}(\Lambda_2+\Lambda_7)\quad\quad \Sigma_3 = \frac{1}{2}(\sqrt{3}\Lambda_8+\Lambda_3)
\ee




The beable for the HV
theory id ${\cal O}={\vec \beta}\cdot{\vec S}$ as before except that now the ${\vec \beta}$ has only three components. But the number of 
parameters required to specify a state is four, and obviously the expectation values of the three $\Sigma_i$ do not suffice to specify
a state. As before we shall use expectation values of 8 independent operators though, as before, for pure states they are functions of 
four real parameters only. For this purpose, it helps to extend the $\Sigma$'s also to eight traceless Hermitian operators. One such set
is provided by
\be
\label{eq:Sigmaextend}
\Sigma_4 = \frac{1}{\sqrt{2}}(\Lambda_1-\Lambda_6)\quad\quad \Sigma_5 = \frac{1}{\sqrt{2}}(\Lambda_2-\Lambda_7)\quad\quad \Sigma_6 = \frac{1}{2}(\sqrt{3}\Lambda_8-\Lambda_3)
\ee
and,
\be
\label{eq:Sigmaextend2}
\Sigma_7 = \Lambda_4\quad\quad \Sigma_8 = \Lambda_5
\ee
The structure constants in the new representation can easily be worked out. It will turn out to be useful to record the explicit matrix
representations for $\Sigma_x^2,\Sigma_y^2,\Sigma_z^2$:
\be
\label{eq:Sigmasquares}
\Sigma_x^2 = \begin{pmatrix}
    \frac{1}{2} & 0 & \frac{1}{2}\\
    0 & 1 & 0\\
    \frac{1}{2} & 0 & \frac{1}{2}
  \end{pmatrix},
\Sigma_y^2 = \begin{pmatrix}
    \frac{1}{2} & 0 & -\frac{1}{2}\\
    0 & 1 & 0\\
    -\frac{1}{2} & 0 & \frac{1}{2}
  \end{pmatrix},
\Sigma_z^2 = \begin{pmatrix}
    1 & 0 & 0\\
   0 & 0 & 0\\
    0 & 0 & 1
  \end{pmatrix} 
\ee
Now we can use any of the four cases that we discussed. Let us, for instance, use case III, the first of the consistent possibilities. Using
$\lambda_1=0,\lambda_2=|{\vec \beta}|,\lambda_3=-|{\vec \beta}|$(it should be appreciated that the identification of $\lambda$'s can be
made in any way), the outcome formula reads
\be
\label{eq:ksbetaoc}
({\vec\beta}\cdot{\vec S})_{OC}\,=\,\frac{|{\vec\beta}|}{2}\,\chi^\beta_1\,(1\,-\,\chi^\beta_2)
\ee
with
\bea
\label{eq:ksbetachis}
\chi^\beta_1&=&\,sign(\mu_1+\frac{1}{2}\,|\frac{p_2-p_3}{1-p_1}|)\,sign(p_2-p_3)\\
\chi^\beta_2&=&\,sign(\mu_2+\frac{1}{2}\,|2p_1-1|)\,sign(2p_1-1)\\
\eea
Our focus is on the BKS theorem so we need to look at the {\it mutually commuting} operators $\Sigma_x^2,\Sigma_y^2,\Sigma_z^2$ and
their associated beables. We shall make use of eqn.(\ref{eq:quadtest}) in this context. Let us illustrate with the example of $(S_x)_{OC}$;
\be
\label{eq:kssxoc}
(S^x)_{OC}\,=\,\frac{1}{2}\,\chi^x_1\,(1\,-\,\chi^x_2)
\ee
This HV outcome rule is not only compatible with the quantum mechanical outcomes of measuring the operator $\Sigma_x$, it also correctly
reproduces the expectation value of $\Sigma_x$ as well as the variance of it. By using the rule of eqn.(\ref{eq:quadtest}),
\be
\label{eq:kssxsqoc}
(S_x^2)_{OC} = ((S^x)_{OC})^2= \frac{1}{2}\,(1\,-\,\chi^x_2)
\ee
On the quantum mechanical side, the mutually commuting $\Sigma_x^2,\Sigma_y^2,\Sigma_z^2$ have simultaneous eigenstates with eigenvalues
adding upto 2 for each operator. Since the operators add up to twice the identity, for a given eigenstate, the eigenvalues of the three
operators also add upto 2. For each simultaneous eigenstate we associate a probability. 
This is summarised in the next table:

\begin{center}
 \begin{tabular}{||c  |c  |c |c |c |c  |} 
 \hline
   $ \Sigma_x^2$ ~ & $ \Sigma_y^2$ ~ &$ \Sigma_z^2$ ~ & Simultaneous Eigenvectors& Probability  \\ 
 \hline\hline
  $1$ & $0$ &$1$ & $ \frac{1}{\sqrt{2}}\begin{pmatrix}
    1\\
    0\\
    1
  \end{pmatrix} $& $p_2$ \\ 
 \hline
  $0$ & $1$ & $1$ & $\frac{1}{\sqrt{2}} \begin{pmatrix}
    1\\
    0\\
    -1
  \end{pmatrix} $& $p_1$ \\ 
 \hline
  $1$ & $1$ &$0$ & $ \begin{pmatrix}
    0\\
    1\\
    0
  \end{pmatrix} $& &$p_3$ \\ 
 \hline
\end{tabular}
\end{center}

\textbf{Hidden Variable Theory}

On the HV side we have the outcomes formulae and care should be exercised in making sure that the correct probabilities enter the
sign functions.

\begin{center}
 \begin{tabular}{||c |c |c |c |c |c |c | c||} 
 \hline
    $(S_x^2)_{OC}$ & $(S_y^2)_{OC}$ & $(S_z^2)_{OC}$ & Probability  \\ [0.5ex] 
 \hline\hline
   ${\lambda_1} = 0$ & $\lambda_3 = 1$ &$\lambda_2 = 1$ & $p_1$ \\ 
 \hline
  $\lambda_2 = 1$ & ${\lambda_1} = 0$ &$\lambda_3 = 1$ & $p_2$ \\ 
 \hline
 $ \lambda_3 = 1$ & $ \lambda_2 = 1$ & $ {\lambda_1} = 0$ &  $p_3$ \\ [1ex]
 \hline
\end{tabular}
\end{center}

%
%
Suppose the the outcome $\lambda_1 = 0$ has probabilities $p_1, p_2, p_3$ for the beables $S_x^2, S_y^2 , S_z^2$ as shown in the above table. 
Thenthe probabilities of the outcomes $\lambda_2 = \lambda_3 = 1$ are $1-p_1, 1-p_2, 1-p_3$ for the beables $S_x^2, S_y^2 , S_z^2$ respectively.
Now from the outcome formulae we have,
\be
(S_x^2)_{OC} = \frac{1}{2}(1 - \chi_2^{\hat{x}})
\ee
Similarly,
\be
(S_y^2)_{OC} = \frac{1}{2}(1 - \chi_2^{\hat{y}})
\ee
\be
(S_z^2)_{OC} = \frac{1}{2}(1 - \chi_2^{\hat{z}})
\ee
Where,
\be
\chi_2^{\hat{x}} = sign(\mu + \frac{1}{2}|2p_1-1|)sign(2p_1-1)
\ee
\be
\chi_2^{\hat{y}} = sign(\mu + \frac{1}{2}|2p_2-1|)sign(2p_2-1)
\ee
\be
\chi_2^{\hat{z}} = sign(\mu + \frac{1}{2}|2p_3-1|)sign(2p_3-1)
\ee

Let us now calculate the expectation of the K-S  beable $B = S_x^2 + S_y^2 + S_z^2$. Taking the expectation values of a sum
of beables to be the sum of the expectation values, and using the results of Sec.3, it follows that
\bea
\label{eq:ksavhv}
\langle B \rangle  &=& \langle S_x^2 \rangle + \langle S_y^2 \rangle + \langle S_z^2 \rangle \nonumber\\ &=& \frac{1}{2}(1-2p_1+1) + \frac{1}{2}(1-2p_2+1) + \frac{1}{2}(1-2p_3+1) \nonumber\\ &=& 2
\eea
This means that on the average the KS-constraint is obeyed by our HV model! But we know that the BKS theorem to be violated. If it is fine
on the average, a violation of BKS would show up through a nontrivial dispersion in B. In order to compute the dispersion in B we need to
compute the average of
\be
\label{eq:kssq}
B^2\,=\,S_x^4+\,S_y^4\,+S_z^4\,+\,2S_x^2\,S_y^2\,+\,2S_x^2\,S_z^2\,+\,S_y^2\,S_z^2
\ee
The average of the fourth power terms can be calculated by successive use of eqn.(\ref{eq:quadtest}). The nontrivial parts are the 
cross terms like $S_x^2 S_y^2$.From the two equations given below,
\be
(S_x)^2_{OC} = \frac{1}{2}(1-\chi_2^{\hat{x}}) = \frac{1}{2} [1 - sign(\mu + \frac{1}{2}|2p_1-1|)sign(2p_1-1)]
\ee
and,
\be
(S_y)^2_{OC} = \frac{1}{2}(1-\chi_2^{\hat{y}}) = \frac{1}{2} [1 - sign(\mu + \frac{1}{2}|2p_2-1|)sign(2p_2-1)]
\ee
The expectation of $S_x^2 S_y^2$ is given by,
\bea
\langle S_x^2 S_y^2 \rangle &=& \frac{1}{4} \int_{-\frac{1}{2}}^{\frac{1}{2}} d\mu [1 - sign(\mu + \frac{1}{2}|2p_1-1|)sign(2p_1-1) 
- sign(\mu + \frac{1}{2}|2p_2-1|)sign(2p_2-1)\nonumber\\ 
&&+ sign(\mu + \frac{1}{2}|2p_1-1|)sign(2p_1-1) sign(\mu + \frac{1}{2}|2p_2-1|)sign(2p_2-1) ] \nonumber\\ 
&=& \frac{1}{4} [1 + (2p_1-1) + (2p_2-1) + (1-||2p_1-1|-|2p_2-1||)sign(2p_1-1)sign(2p_2-1)]\\
\eea
Therefore,
\bea 
\langle (S_x^2 + S_y^2 + S_z^2)^2 \rangle &=& \langle S_x^4  \rangle + \langle S_y^4  \rangle + \langle S_z^4  \rangle 
+ 2 [\langle S_x^2 S_y^2  \rangle + \langle S_y^2 S_z^2  \rangle + \langle S_z^2 S_x^2  \rangle] \nonumber\\
&=& (p_1-1) + (p_2-1) + (p_3-1) \nonumber\\ && + \frac{1}{2} [5 + \sum_{i<j}(1-||2p_i-1|-|2p_j-1||)sign(2p_i-1)sign(2p_j-1)] \nonumber\\ 
&=& 4 + \frac{1}{2} [1 + \sum_{i<j}(1-||2p_i-1|-|2p_j-1||)sign(2p_i-1)sign(2p_j-1)]\\
\eea
Thus generically, the dispersion is nonvanishing. There are two implications of that: i. configuration by configuration the K-S constraint
is not satisfied which is a contradiction with the theorem. ii. Since the dispersion in quantum theory vanishes, any dispersion at the HV 
level contradicts quantum mechanics and we do not have an acceptable hidden variable theory.

It is instructive to get an idea of how large the dispersion can be. Note that all cases can be divided into
Only two possible cases:

\textbf{Case I :} When $p_1, p_2, p_3 < \frac{1}{2}$, all $sign(2p_i-1)$ are \textbf{positive} 

\textbf{Case II :} When $p_1 > \frac{1}{2}; p_2, p_3 < \frac{1}{2}$, two $sign(2p_i-1)$ are \textbf{positive} and one is \textbf{negative}

\textbf{Examples}

\textbf{1.} For $p_1 = p_2 = 0; p_3 = 1$ , $\langle (S_x^2 + S_y^2 + S_z^2)^2 \rangle = 4$ , which is the \textbf{Minimum}

\textbf{2.} For $p_1 = p_2 = p_3 = \frac{1}{3}$ , $\langle (S_x^2 + S_y^2 + S_z^2)^2 \rangle = 6$ , which is the  \textbf{Maximum}

Thus the maximum variance in the K-S operator is 2 and the minimum is 0. This last possibility is intriguing.

\subsection{Another approach}

Now we show that it is possible to construct HV models where the dispersion in the K-S beable can be made arbitrarily small but never
zero. But the price to be paid by such models is that generically they will fail to reproduce even the average values of linear beables.
Let us consider the beable $B_\epsilon\,=\,(1+\epsilon) S_x^2 + S_y^2 +(1- \epsilon) S_z^2$

The following table summarises the quantum mechanical aspects of the problem.
\begin{center}
 \begin{tabular}{||c |c |c |c |c |c |c | c||} 
 \hline
   $((1+\epsilon) \Sigma_x^2 + \Sigma_y^2 + (1-\epsilon) \Sigma_z^2)_{OC}$  &  Eigenvectors  \\ [0.5ex] 
 \hline\hline
  $ 2  $ & $ \frac{1}{\sqrt{2}}\begin{pmatrix}
    1\\
    0\\
    1
  \end{pmatrix}$  \\ 
 \hline
  $ 2-\epsilon $ & $\frac{1}{\sqrt{2}} \begin{pmatrix}
    1\\
    0\\
    -1
  \end{pmatrix} $ \\ 
 \hline
  $2+\epsilon$ & $ \begin{pmatrix}
    0\\
    1\\
    0
  \end{pmatrix} $ \\ [1ex]
 \hline
\end{tabular}
\end{center}
The HV description is summarised in the table below. 
\begin{center}
 \begin{tabular}{||c |c |c |c |c |c |c | c||} 
 \hline
    $((1-\epsilon) S_x^2 + S_y^2 + (1-\epsilon) S_z^2)_{OC}$ & Probability  \\ [0.5ex] 
 \hline\hline
  $ 2  $ & $p_0$ \\ 
 \hline
  $ 2-\epsilon $ & $p_-$ \\ 
 \hline
 $2+\epsilon$ &  $p_+$ \\ [1ex]
 \hline
\end{tabular}
\end{center}
The probabilities satisfy $p_++p_0+p_-=1$. We can apply our general results for
three arbitrary eigenvalues to obtain the various outcome formulae. From our general results it is clear that 
the HV description so constructed will agree with quantum mechanics for both the average and variance of $B_\epsilon$.
We shall however bypass all those details by calculating these in terms of the probabilities. However, to clarify an
important point we shall just give the outcome formula for $S_x^2$; similar considerations hold for the other beables in question.

%
%
%
%
%
%
\be
\label{eq:squareoc}
{( S_x^2)}_{OC} = \frac{3}{4} - \frac{1}{4} \chi_1 +  \frac{1}{4} \chi_2 +  \frac{1}{4} \chi_1 \chi_2
\ee
Now, the expectation value of $B_\epsilon$ is given by
\be
\label{eq:bepsav}
\langle\,B_\epsilon\,\rangle_{HV} = p_+(2+\epsilon)\,+\,p_0\,\cdot 2\,+\,p_-(2-\epsilon)\,=\,2\,+\,\epsilon(p_+-p_-)
\ee
Likewise, the average of $B_\epsilon^2$ is given by
\be
\label{eq:bepsqav}
\langle\,B_\epsilon^2\,\rangle_{HV}\,=\,4\,+\,4\epsilon(p_+-p_-)\,+\,\epsilon^2\,(p_++p_-)
\ee
Hence the variance is given by
\be
\label{eq:bepsvar}
\Delta^2\,=\,\epsilon^2\,(p_++p_-\,-\,(p_++p_-)^2)
\ee
Thus the average can be made to approach 2, and the variance 0, arbitrarily closely. But the catch now is in predicting the averages
and variances of linear operators in accordance with quantum mechanics. The essential difficulty is there is no way to go from outcome
formulae for squares of the type in eqn.(\ref{eq:squareoc}) to outcome formulae for linear operators without introducing additional
sign functions whose effect generically is to make the averages for linear operators disagree with quantum mechanics. Consequently HV theories
of this type are not acceptable.

\section{von Neumann's homogeneity aspect in Bell's HV prescription}

von Neumann had an independent, and stronger, objection to HV theories which he paraphrased through the  question:i ``Is it possible to have 
homogeneous ensembles even if there is no dispersion-free ensemble ?" Homogeneous ensembles are those whose each and every subensemble gives 
rise to the same expectation value which is same as the expectation value for the whole ensemble itself.
The interpretation of this aspect of von Neumann's argument is that the necessarily inhomogeneous ensembles of HV theory parametrised by 
$\epsilon_i$ over and above the hidden variable
$\lambda$'s, distributed, in the above example, uniformly, can not truly {\bf mimic} a {\bf homogeneous} quantum ensembles of states.

But Bell's prescription of hidden variable formula is directly violating the homogeneity of the ensemble over the hidden variable $\lambda$. 
Suppose for $\sum_i\,\beta_i\,\epsilon_i > 0$ we divide the ensemble into two parts over $\lambda$ as follows:
{\bf $\lambda_+$:} $ -\frac{1}{2|\vec{\beta}|}|\sum_i\,\beta_i\,\epsilon_i| < \lambda < \frac{1}{2}$; {\bf $\lambda_-$}:
$ -\frac{1}{2} < \lambda < -\frac{1}{2|\vec{\beta}|}|\sum_i\,\beta_i\,\epsilon_i|$.

Then, 
\bea
\label{eq:homoplus}
\langle \alpha + \vec{\beta}\cdot\vec{S} \rangle_{\lambda_+}& =& \int_{-\frac{1}{2|\vec{\beta}|}|\sum_i\,\beta_i\,\epsilon_i|}^{1/2} 
\frac{d\lambda}{\frac{1}{2}+\frac{1}{2|\vec{\beta}|}|\sum_i\,\beta_i\,\epsilon_i|} (\alpha + \vec{\beta}\cdot\vec{S})_{OC} \nonumber\\
&=& \frac{1}{\frac{1}{2}+\frac{1}{2|\vec{\beta}|}|\sum_i\,\beta_i\,\epsilon_i|} (\alpha + |\vec{\beta}|)
(\frac{1}{2}+\frac{1}{2|\vec{\beta}|}|\sum_i\,\beta_i\,\epsilon_i|)\nonumber\\
&=& \alpha + |\vec{\beta}|
\eea
Likewise,
\bea
\label{eq:homominus}
\langle \alpha + \vec{\beta}\cdot\vec{S} \rangle_{\lambda_-}& =& \int_{-1/2}^{-\frac{1}{2|\vec{\beta}|}|\sum_i\,\beta_i\,\epsilon_i|} 
\frac{d\lambda}{\frac{1}{2}-\frac{1}{2|\vec{\beta}|}|\sum_i\,\beta_i\,\epsilon_i|} (\alpha + \vec{\beta}\cdot\vec{S})_{OC}\nonumber\\
&=& \frac{1}{\frac{1}{2}-\frac{1}{2|\vec{\beta}|}|\sum_i\,\beta_i\,\epsilon_i|}(\alpha - |\vec{\beta}|)(\frac{1}{2}-\frac{1}{2|\vec{\beta}|}|
\sum_i\,\beta_i\,\epsilon_i|)\nonumber\\
&=& \alpha - |\vec{\beta}|
\eea
And certainly, $\alpha + |\vec{\beta}| \neq \alpha - |\vec{\beta}|$.
Hence Bell-like prescriptions can not really mimic a \emph{homogeneous quantum ensemble}. 
 But the irony is that this is very difficult to observe this experimentally because if we choose
the subensembles ito be sufficiently large, the $\lambda$-distribution within them will be practically uniform.

\section{Results}
We simplifed Bell's construction for spin-$\frac{1}{2}$ particles to make things more transparent. His result that the expectation 
values of beables agrees exactly with the predictions of Quantum Mechanics was reproduced by our simplified treatment.

Since an acceptable Hidden Variable Theory must reproduce all the predictions of quantum mechanics, Bell's example should do so too. Not 
only averages, but dispersions must be reproduced too. 
Since in the case of spin-$\frac{1}{2}$, the outcomes of the HV theory obey the same algebra as the
operators of quantum theory i.e their squares are identity, dispersions are trivially reproduced. 
Additionally, other features like outcomes of sequential measurements on a single 
copy must tally with quantum mechanics too. 
This can happen in HV theories only if measurements deterministically change states.
But no explicit models of measurements are to be
found that reproduce the quantum mechanical results, though such claims are made in Bohmian Mechanics.

We then discuss how Bell's example directly violates the homogeneity of the corresponding quantum ensemble. In particular, we show that though 
it is in principle possible to empirically distinguish between HV theories and quantum mechanics, it is practically impossible to do so.

Motivated by the desire to find examples where dispersions are not trivially reproduced, we turned to finding a Bell-like example for
spin-1 systems.
It is well known that due to Bell-Kochen-Specker theorems one can not construct acceptable HV theories of spin-1. 
Nevertheless we produce a Hidden Variable formula of Bell's kind for spin-1 which reproduces exactly both the expectation values
of arbitrary beables(involving all the \emph{eight} independent observables of the theory) as well as their dispersions as predicted by 
Quantum Mechanics. We have shown how they explicitly violate the BKS theorem. 

To further understand the implications of the Kochen-Specker-Bell theorems, we constructed a variant where the beables are the three 
angular momentum
operators. It too reproduces the expectation values and dispersions. Though state by state it violates these theorems, the averages of the
so called K-S constraints are exactly reproduced. We constructed another HV theory that violates BKS arbitrarily little but it is at
variance with other predictions of quantum mechanics.

In summary, Bell-like examples of HV theories can not always be considered as counterexamples to von Neumann unless they are shown to
reproduce all the features of quantum mechanics.

%

\section*{Acknowledgements}
One of us, Kaushik Borah, wishes to thank to the DST-Inspire, Govt. of India and the TIFR-TCIS, Hyderabad for giving him the opportunity 
to have a useful time in the summer. He also wishes to thank to Prof. Subroto Mukerjee, IISc, for some useful discussions.

\end{document}